# Inflation, unemployment, and labour force. Phillips curves and long-term projections for Austria

Ivan Kitov, Institute for the Geospheres' Dynamics, Russian Academy of Science

**Abstract**

We model the rate of inflation and unemployment in Austria since the early 1960s within the Phillips/Fisher framework. The change in labour force is the driving force representing economic activity in the Phillips curve. For Austria, this macroeconomic variable was first tested as a predictor of inflation and unemployment in 2005 with the involved time series ended in 2003. Here we extend all series by nine new readings available since 2003 and re-estimate the previously estimated relationships between inflation, unemployment, and labour force. As before, a structural break is allowed in these relationships, which is related to numerous changes in definitions in the 1980s. The break year is estimated together with other model parameters by the Boundary Element Method with the LSQ fitting between observed and predicted integral curves. The precision of inflation prediction, as described by the root-mean-square (forecasting) error is by 20% to 70% better than that estimated by AR(1) model. The estimates of model forecasting error are available for those time series where the change in labour force leads by one (the GDP deflator) or two (CPI) years. For the whole period between 1965 and 2012 as well as for the intervals before and after the structural break (1986 for all inflation models) separately, our model is superior to the naïve forecasting, which in turn, is not worse than any other forecasting model. The level of statistical reliability and the predictive power of the link between inflation and labour force imply that the National Bank of Austria does not control inflation and unemployment beyond revisions to definitions. The labour force projection provided by Statistic Austria allows foreseeing inflation at a forty-year horizon: the rate of CPI inflation will hover around 1.3% and the GDP deflator will likely sink below zero between 2018 and 2034.

**Introduction**

Price stability is a key responsibility of central banks. Following the Treaty on the Functioning of the European Union, the National Bank of Austria (Oesterreichische Nationalbank, OeNB) has established a quantitative definition of price stability, which is below but close to 2% (OeNB, 2013). Monetary policy has to provide the target value, and thus, should be supported by an extensive analytical framework designed for decision-making process. Within this framework, economic analysis focuses on output and prices. The comprehension of driving force(s) behind the short-term fluctuations and the long-term evolution of price inflation and unemployment creates a solid basis for a sound monetary policy.

The literature on inflation in Austria is not as extensive as for the USA and chiefly belongs to the authors from the Austrian Institute of Economic Research (WIFO) (*e.g.*, Baumgartner, 2002), the Institute for Advanced Studies (IHS) (*e.g.*, Hofer and Koman, 1991), and the OeNB (*e.g.*, Fritzer, 2011; Fritzer *et al.*, 2008). There are a few studies comparing inflation models developed by all three institutions (*e.g.*, Ragacs and Schneider, 2007). Considering the quality of data measured since the early 1950s and the accuracy of econometric modelling the Austrian economy deserves to be in the focus of professional attention of the broader economic community.

Two economic crises in the 21$^{st}$ century have significantly reshaped the set of tools for description and prediction of inflation in developed countries putting forward the concept of New Keynesian Phillips Curve (NKPC). Galí and Gertler (1999) formulated a quantitative model for "inflation expectations" controlled (or anchored) by various instruments of monetary and fiscal policy. This concept has placed central banks in focus of a comparative inflation study related to the differences in monetary policy and its results in developed countries (Galí *et al.*, 2001; Sims, 2007). Galí and Monacelli (2005) introduced the case of small open economies with terms of trade extending the original set of defining parameters, which was considered as the most relevant to the Austrian economy. However, Rumler (2007) and Mikhailov *et al.* (2008) modelled the inflation dynamics in small open economies using the NKPC and found just a moderate support to the terms of trade (external factor) as a driver of price inflation in Austria.

Kitov (2006) constructed a Phillips curve (PC) for Austria as a closed economy using the change in labour force as the only driver of inflation and unemployment. The modelling error for the period between 1965 and 2003 with a structural break near 1986 was smaller than that for any other structural model for the same period. The change in labour force explained 81% of the variability ($R^2$=0.81) in the rate of CPI inflation between 1965 and 2003 with the root-mean-square error (RMSE) for the best fit model of ~0.01 y$^{-1}$ (1% per year). In this study, we extend the time series to 2012, revisit the original model, and validate it. Our results demonstrate that the OeNB hardly controls inflation and unemployment, but relies on the slow change in the workforce since the late 1990s.

Forecasting is an important part of economic analysis for the purposes of central banks. Rumler and Valderrama (2010) forecasted inflation using a single-equation NKPC and systematically compared their results to forecasts generated from a traditional Phillips Curve, a Bayesian VAR (which is also used for the first time to forecast Austrian inflation), a conventional VAR, an AR model and the naive forecast for 1-quarter, 4-quarters and 8-quarters. They found that the NKPC beats the forecasts derived from the time series models, the traditional PC, and the naive forecast in terms of lower RMSE only for longer forecast horizons of 1 and 2 years. The use of labour force as the driving force of inflation (*e.g.*, Kitov and Kitov, 2010) converts inflation forecasting into labour force projection. Then, the short-term (a few months) inflation forecasts might be spoiled by noisy data from labour force surveys and CPI estimates, but the mid- and long-term projections of working age population and labour force participation models are able to accurately predict the evolution of prices treated as "inflation expectations" in the NKPC models.

The remainder of this paper consists of two Sections and Conclusion. Section 1 briefs on major developments within the Phillips curve framework, introduces a set of linear and lagged relationships between studied parameters, and describes the Boundary Element Method used to estimate coefficients in these relationships. Section 2 presents a series of revised inflation and unemployment models for Austria and reports some quantitative/statistical results for two individual and the generalized link between labour force, inflation, and unemployment.

## 1. The Fisher/Phillips curve framework

Irving Fisher (1926) introduced price inflation as driving the rate of unemployment. He modelled monthly data between 1915 and 1925 using inflation lags up to five months. The inflation and unemployment time series were short and contained higher measurement errors to produce robust statistical estimates of coefficients and lags in the relevant causal relationship. Kitov (2009) estimated a Fisher-style relationship for the USA using observations between 1965 and 2008 and found that the change in unemployment lags behind the change in inflation by 10 quarters. The 43-year period provides good resolution and high statistical reliability of both regression coefficients and the lag. This relationship was successfully tested for cointegration and Granger causality. The two-and-a-half year lag implies the only order of occurrence. But other countries may demonstrate different lags and order (Kitov and Kitov, 2010).

Phillips (1958) interpreted the link between (wage) inflation and unemployment in the UK in the opposite direction. The original Phillips curve implied a causal and nonlinear link between the rate of change of the nominal wage rate and the contemporary rate of unemployment. He suggested that wages are driven by the change in unemployment rate. The assumption of a causal link worked well for some periods in the UK. When applied to inflation and unemployment measurements in the USA, the PC successfully explained the 1950s. Then, the PC became an indispensible part of macroeconomics which has been extensively used by central banks ever since. The success of the PC did not last long, however, and new data measured in the late 1960s and early 1970 challenged the original version. When modelling inflation and unemployment in Austria, we follow up the original assumption of a causal link between inflation and unemployment to construct an empirical Fisher/Phillips-style curve.

The period of fast inflation growth in the late 1960s and 1970s brought significant changes to the original PC concept. The mainstream theory had to include autoregressive properties of inflation and unemployment in order to explain the observations. For the sake of quantitative precision, the rate of unemployment was replaced by different parameters of economic and financial activity. All in all, the underlying assumption of a causal link between inflation and unemployment was abandoned and replaced by the hypothesis of "rational expectations" (Lucas, 1972, 1973), and later by the concept of "inflation expectations" (Galí and Gertler, 1999). The former approach includes a varying number of past inflation values (autoregressive terms). It was designed to explain inflation persistency during the high-inflation period started in the early 1970s and ended in the mid 1980s.

The concept of inflation expectations surfaced in the late 1990s in order to explain the Great Moderation (Clarida *et al.*, 2000; Cecchetti *et al.*, 2004; Bernanke, 2004) as controlled by monetary and fiscal authorities (Sims, 2007, 2008). The term "New Keynesian Phillips Curve" was introduced in order to bridge this new approach to the original Keynesian framework (Gordon, 2009). The number of defining parameters has dramatically increased in the NKPC (a few autoregressive terms with varying coefficients) relative to the parsimonious Phillips curve. However, both approaches have not been successful in quantitative explanation and prediction of inflation and/or unemployment (*e.g.*, Rudd and Whelan, 2005ab).

Stock and Watson (1999) were outspoken on data and tested a large number of Phillips-curve-based models for predictive power using various parameters of activity (individually and in aggregated form) instead of and together with unemployment. This purely econometric approach did not include extended economic speculations and was aimed at finding technically appropriate predictors. The principal component analysis (Stock and Watson, 2002) was a natural extension to the multi-predictor models and practically ignored any theoretical background. Under the principal component approach, the driving forces of inflation are essentially hidden.

The original Phillips curve for the UK and the Fisher curve, which could be named as an "anti-Phillips curve", both provide solid evidences for the existence of a causal link between inflation and unemployment. The conflict between the directions of causation can be resolved when both variables are driven by a third force with different lags. Depending on which lag is larger inflation may lag behind or lead unemployment. Co-movement found in Japan is just a degenerate case (Kitov and Kitov, 2010).

The framework of our study is similar to that introduced and then developed by Stock and Watson (2006, 2007, 2008) for many predictors. They assessed the performance of inflation forecasting in various specifications of the Phillips curve. Their study was forced by the superior forecasting result of a univariate model (naïve prediction) demonstrated by Atkeson and Ohanian (2001). Stock and Watson convincingly demonstrated that neither before the 2007 crisis (2007) nor after the crisis (2010) can the Phillips curve specifications provide long term improvement on the naïve prediction at a one-year horizon.

Following Fisher and Phillips, we do not include autoregressive components in the Phillips curve and estimate two different specifications for inflation:

$$\pi(t) = \alpha + \beta u(t-t_0) + \varepsilon(t) \quad (1)$$

$$\pi(t) = \alpha_1 + \beta_1 l(t-t_1) + \varepsilon_1(t) \quad (2)$$

where $\pi(t)$ is the rate of price inflation at time $t$, $\alpha$ and $\beta$ are empirical coefficients of the Phillips curve with the time lag $t_0$, which can be positive or negative, $u(t)$ is the rate of unemployment, and $\varepsilon(t)$ is the error term, which we minimize by the least squares (LSQ) method applied to the cumulative curves, with the initial and final levels fixed to the observed ones. In (2), $l(t)=dlnLF(t)/dt$ is the rate of change in labour force, $\alpha_1$ and $\beta_1$ are empirical coefficients of the link between inflation and labour force, $t_1$ is the non-negative time lag of inflation, and $\varepsilon_1(t)$ is the model residual.

Then, we represent unemployment as a linear and lagged function of the change rate in labour force:

$$u(t) = \alpha_2 + \beta_2 l(t-t_2) + \varepsilon_2(t) \quad (3)$$

with the same meaning of the coefficients and the lag as in (2). We finalize the set of causal models with a generalized version:

$$\pi(t) = \alpha_3 + \beta_3 l(t-t_1) + \gamma_3 u(t-t_0) + \varepsilon_3(t) \quad (4)$$

Relationships (2) through (4) have been re-estimated with the data for the past nine years and the Boundary Element Method (BEM) instead of standard regression.

The BEM converts linear (also partial) differential equations, *e.g.* relationships (2) through (4), to a set of integral equations. The solution of the integral equations for the period between $t_0$ and $t_{01}$ is an exact solution of the original differential equations. For relationship (2):

$$\int_{t_0}^{t_{01}} d[lnP(t)] = \int_{\tau_0}^{\tau_{01}} (\beta_1 d[lnLF(\tau)] + \int_{\tau_0}^{\tau_{01}} \alpha_1 d\tau + \int_{\tau_0}^{\tau_{01}} \varepsilon_1(\tau) d\tau \quad (5)$$

where $\pi(t)$ is by the rate of change in the price level, $P(t)$, $\tau=t-t_1$, and $\int_{\tau_0}^{\tau_{01}} \varepsilon_1(\tau) d\tau = 0$. The solution of the integral equation (5) is as follows:

$$lnP \big|_{t_0}^{t_{01}} = \beta_1 lnLF \big|_{\tau_0}^{\tau_{01}} + \alpha_1 t \big|_{\tau_0}^{\tau_{01}} + C \quad (6)$$

where $C$ is the free term ($C=0$), which has to be determined together with coefficients $\alpha_1$ and $\beta_1$ from the boundary conditions: $P(t_0)=P_0$, $P(t_{01})=P_1$, $LF(\tau_0)=LF_0$, and $LF(\tau_{01})=LF_1$. For 1-D problems, we have fixed values as boundary conditions instead of boundary integrals. The number of boundary conditions in (6) is complete for calculation (or quantitative estimation, if there is no analytic solution) of all involved coefficients. Without loss of generality, one can always set $P_0=1.0$ as a boundary condition. The estimated coefficients $α_1*$, $β_1*$, and $C*$ entirely define the particular solution of (6):

$$ln[P(t_{01})] = β_1*ln[LF(τ_0)/LF(τ_{01})] + α_1*(τ_{01}-τ_0) \qquad (7)$$

at $t_{01}$, as well as over the entire time interval between $t_0$ and $t_{01}$. It is presumed that $LF(t)$ is a discrete function known from measurements.

The estimation of all involved coefficients gives numerical solutions of 2-D and 3-D problems by the BEM in scientific and engineering applications. In this study, the least-square method is used to estimate the best fir coefficients. Therefore, the residual between observed and predicted curves is minimized in the L2 metrics. For solving problem (7) with an increasing accuracy, one can run over a series of boundary conditions for subsequent years.

In terms of the boundary elements method, the right hand side of (7) is the particular solution of the (ordinary) differential equation (2). Since $t_1≥0$, the causality principle holds, and the independent function is known before the dependent one. The only principal difference with the standard BEM used in scientific applications is that the solution (7) is not a closed-form or an analytic solution. The solution is the change in labour force in a given country, which may follow a quite exotic trajectory as related to demographic, social, economic, cultural, climatic, etc. circumstances. From (7), inflation can be exactly predicted at a time horizon $t_1$ and foreseen at longer horizons with various projections of labour force.

In (7), a linear combination of $ln[(LF(t)/LF(t_0))]$ and $(t-t_0)$ defines any particular solution of (2). The rate of price inflation may change only due to the change in labour force. However, the overall price level may grow even when workforce is constant because of $α_1(τ_{01}-τ_0)$ term, for $α_1≠0$.

## 2. Inflation and unemployment models

In terms of working age population, Austria represents an example of a small economy. The Austrian economy has a long history of measurements with open access to all time series and descriptive information. Essentially, the data quality is high. It is the most important characteristic defining the success of any quantitative modelling. We distinguish two main sources of uncertainty related to the data. One source is associated with measurement errors. Due to limited population coverage, the accuracy of labour force surveys is low. Therefore, the original annual figures for unemployment and labour force are not precise. In labour surveys, the measurement accuracy depends on sampling and non-sampling errors. The former is estimated from the population coverage and standard statistical procedures, and the latter is more difficult to evaluate.

Another source of quantitative uncertainty is important for both labour force and inflation measurements. It is associated with the revisions to definitions. In many cases, these revisions are significant, as one can judge from the description given by the OECD (2005). When applied to labour force, the definitional revisions introduce artificial breaks (jumps) in time series as associated with the change in units of measurements. European countries have implemented these revisions at different times creating asynchronous breaks. Some modifications of methodologies and procedures related to inflation measurements are accompanied by the introduction of new measures such as harmonized index of consumer prices (Eurostat, 2013). This index includes rural consumers, but excludes the imputed rent components. It replaced the old CPI definition in the official statistics.

We use six independent sources providing annual readings of CPI, GDP deflator, population estimates, unemployment rate, participation rate, and labour force level: Eurostat, OECD, AMS (Arbeitsmarktservice) Österreich (http://www.ams.or.at), HSV (Hauptverband der Sozialversicherungtraeger) Österreich (http://www.hsv.or.at), Statistik Austria (SA -

http://www.statistik.at), and the Österreichische Nationalbank (ÖNB – http://www.oenb.at). These sources estimate the same variables in different ways. Comparison of formally equivalent time series allows quantitative evaluation of the differences between them. The cross-examination has two main purposes. Firstly, it demonstrates the discrepancy between these series as a quantitative measure of the uncertainty in corresponding parameters. Secondly, we determine the degree of similarity (cross correlation) between these series in order to assess the performance of some true time series, which represent actual values of measured parameters according to perfect definitions.

Data uncertainty puts a strong constraint on the level of confidence related to statistical estimates. One cannot trust statistical inferences with a confidence level higher than that allowed by the intrinsic data uncertainty. On the other hand, equivalent time series obtained according to various definitions (procedures, methodologies, samples, etc.) of the same parameter represent different proportions of the true value for a given parameter. For example, various definitions of employment are aimed at obtaining the number of persons who work for pay or profit, but the minimum level of payment may differ across definitions.

Several definitions are designed to approach the (virtual) true value. If consistent and successful, these definitions may provide estimates, which describe the same portion of the true values over time. Therefore, these successful estimates are scalable - one can easily compute values according to all definitions having only one of them and relevant conversion factors. In that sense, various definitions (and related estimates) should be interchangeable when used to model the link between inflation, unemployment, and labour force.

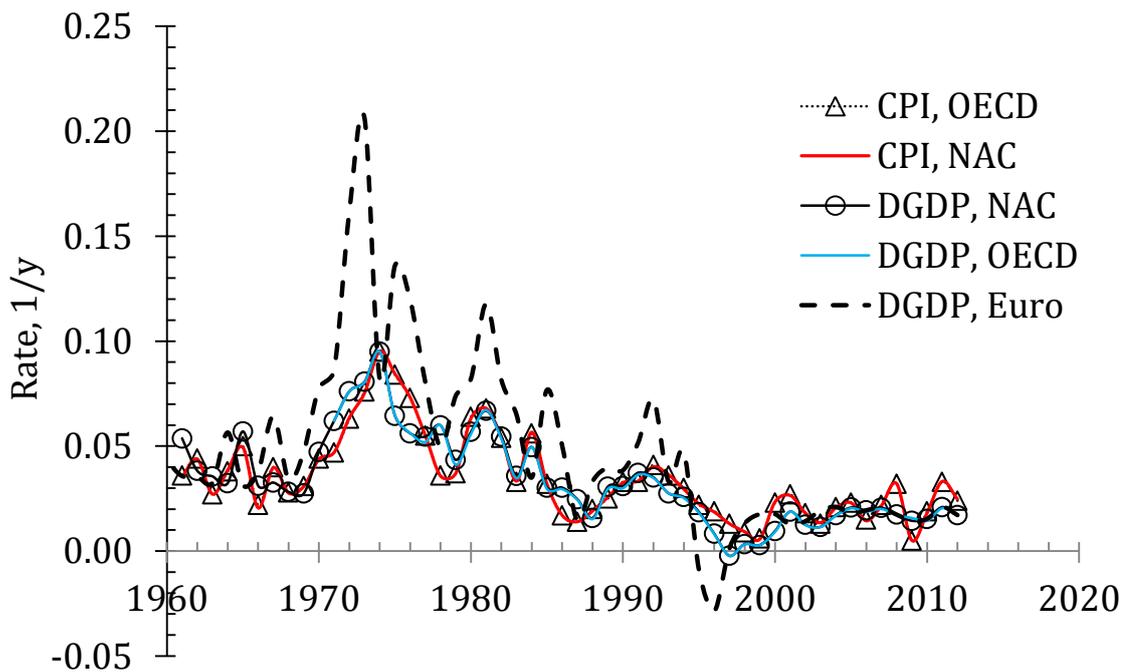

Figure 1. Comparison of five variables representing inflation in Austria.

Five different time series of price inflation, constructed with the use of varying definitions, are presented in Figure 1: CPI and GDP deflator (DGDP) as obtained using prices expressed in national currency (national accounts - NAC), and the GDP deflator estimated using the Austrian shilling/Euro exchange rate before the introduction of Euro. The latter variable is characterized by the largest variations. Two curves representing the NAC CPI and NAC GDP deflator are closer, correlation coefficient (*R*) of 0.91 for the period between 1961 and 2012, but differ in amplitude and timing of principal changes. There are periods of almost total agreement, however. The Euro GDP deflator series is characterized by correlation coefficients of 0.81 and 0.80 as obtained for the NAC DGDP deflator and CPI, respectively. Therefore, we can expect a better interchangeability between the NAC CPI and NAC GDP deflator than that in the other two possible combinations. Finally, the NAC and OECD CPI estimates are almost identical: $R=0.9997$. Since the mid-1970s, the rate of price inflation in Austria has a definition-independent tendency to decrease.

Regarding time incompatibility in the inflation time series, Rumler and Valderrama (2008) mentioned a possibility of a structural break at the end of the 1980s in the inflation data used for the estimation of a Phillips curve for Austria. They reported the absence of statistical evidence of a structural break at any later date in Austria. Mikhailov *et al.* (2008) fixed a break in Austrian economic data to 1991 due to its close economic links with Germany.

In Europe, labour force surveys generally cover small portions of the total population. The levels of labour force and unemployment are estimated using specific weights (population controls) for every person in a given survey to compute the proportion of population with the same characteristics. The population controls in predefined age-sex-race bins are primarily obtained during censuses, which cover the entire population. Between censuses, the population controls are obtained by the population components change: births, deaths, and net migration.

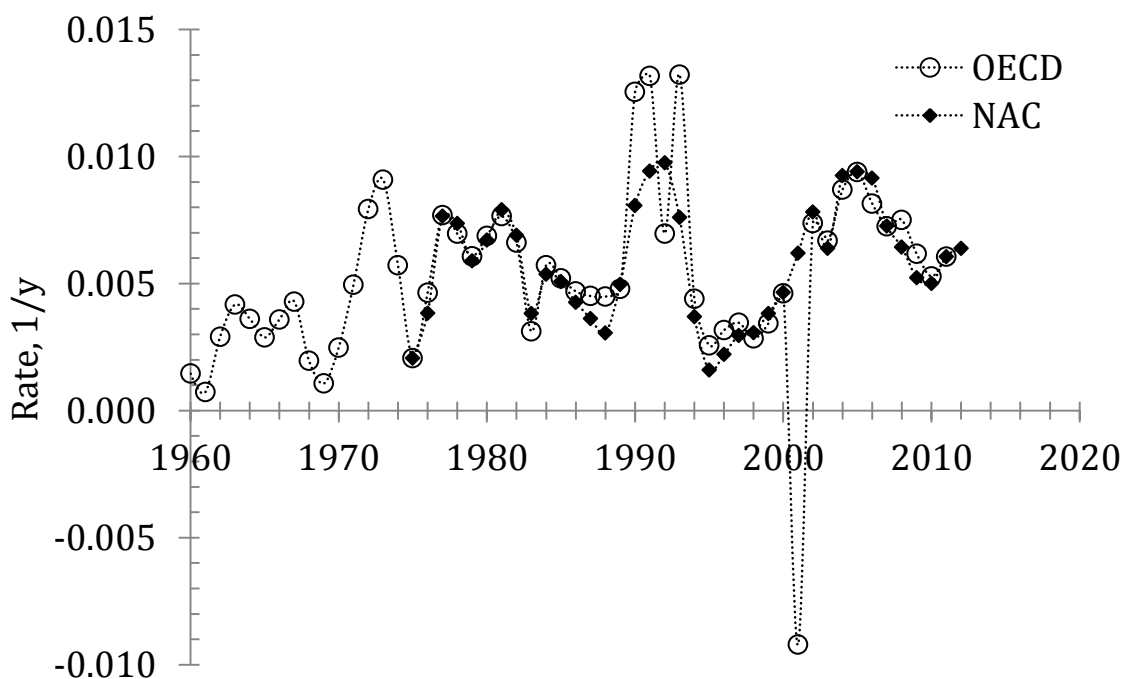

Figure 2. Comparison of the rate of change in working age population (aged 15 and over) as determined by the OECD and national statistics (NAC).

Later revisions to population estimates sometimes reach several percent. Thus, when using current figures of labour force and unemployment, one has to bear in mind that they are subject to further revisions. Figure 2 illustrates the differences in population estimates between the OECD and Statistik Austria (NAC): two curves represent the rate of change in the population of 15 years of age and over. Between 1960 and 1983, the curves coincide since the OECD uses the national definition. After 1983, the curves diverge. There are three distinct peaks in the OECD curve: between 1990 and 1993 and in 2002, which are related to the population revisions. As explained by the OECD (2006) for the population series:

***Series Breaks.*** *From 1992, data are annual averages. Prior to 1992, data are mid-year estimates obtained by averaging official estimates at 31 December for two consecutive year ... From 2002, data are in line with the 2001 census.*

The 2002 revision completely compensates the difference between the OECD and Statistik Austria as accumulated during the previous 20 years: the populations in 1982 and 2002 coincide. Such step adjustments are observed in the USA population data as well. These jumps significantly deteriorate any statistical estimates, but can be completely removed when evenly redistributed over the previous period. Sometimes step adjustments are confused with actual changes in economic variables. One has to be careful in distinguishing between genuine changes and artificial corrections usually associated with the years of census or significant revisions to definitions.

The national estimates in Figure 2 are visually smoother indicating some measures applied to distribute the errors of the closure and other adjustments over the entire period. On average, the population over 15 years of age in Austria has been changing slowly so far – at an annual rate below 0.5% - with a few spikes to the levels of 0.7% to 1.0%. This slow but steady growth supports the gradual increase in labour force.

The rate of labour force growth was low in Austria during the past 15 years, as Figure 3 depicts. Two time series estimated by the OECD and NAC describe the change in labour force. The NAC readings include the estimates of employment made according to the HSV definition and the reports on unemployment made by the AMS. Both agencies base their estimates on administrative records. Thus, their approach has been undergoing weaker changes in definitions and procedures since the 1960s compared to that adopted by the OECD.

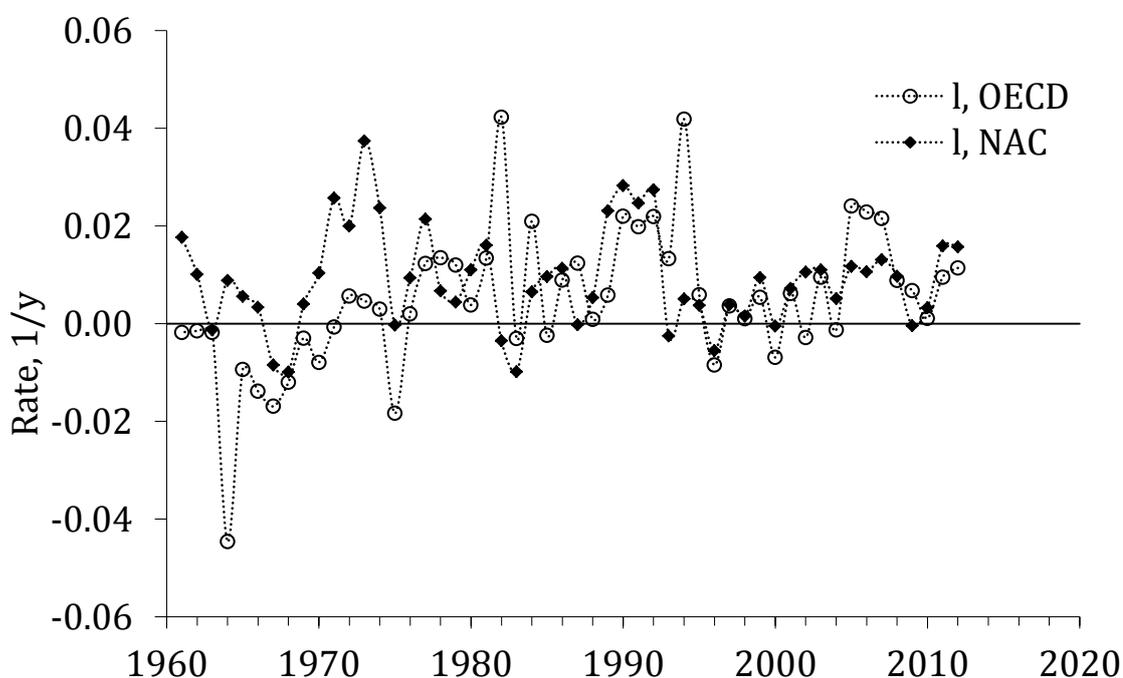

Figure 3. Comparison of the change rate of labour force, as reported by the OECD and NAC.

The curves in Figure 3 inherit the features, which are demonstrated by corresponding working age populations in Figure 2. The OECD curve is characterized by several spikes of artificial character. The NAC curve is smoother. It demonstrates a period of slow growth with high volatility in the late 1970s, a period with an elevated growth rate with high volatility between 1981 and 1995, and again a slow growth period with low volatility during the years from 1995 to 2012. The second period is characterized by significant changes in the labour force definition - both for employment and unemployment (OECD, 2005):

***Series breaks****. In 1982, re-weighting of the sample was made, due to an underestimation of persons aged 15 to 29 years. In 1984, the sample was revised and a change occurred in the classification of women on maternity leave: they were classified as unemployed before 1984 and as employed thereafter. In 1987, a change occurred in the definition of the unemployed where non-registered jobseekers were classified as unemployed if they had been seeking work in the last four weeks and if they were available for work within four weeks. In previous surveys, the* *unemployment concept excluded most unemployed persons not previously employed and most persons re-entering the labour market.*
*Employment data from 1994 are compatible with ILO guidelines and the time criterion applied to classify persons as employed is reduced to 1 hour.*

Therefore, one can expect some changes in the units of labour force measurements during the period between 1982 and 1987 as well as in 1994. The latter change is potentially the largest since the time criterion dropped from 13 hours, as had been defined in 1974, to 1 hour. For the sake of consistency in definitions and procedures, the AMS labour force is used as an independent variable. The OECD labour force is also used in a few cases to illustrate that both definitions provide similar results.

The AMS time series of change rate of labour force contains 62 readings between 1951 and 2012. Here, only the period after 1960 is analyzed and we use 52 readings after 1961 to test for unit roots. This is the first time series we test for stationarity because it is used as a predictor. The inflation time series are likely biased by many revisions to definitions during the modelled period. We discuss unit root tests in the inflation series within the framework of the link between inflation and labour force. The null hypothesis of a unit root in the AMS time series was rejected by the ADF test is -4.22, with 1% critical value of -3.58). For the OECD series, the ADF test is -4.72. The Phillips-Perron test has also rejected the null of a unit root in both series. For the AMS series: $z(\rho)$=-26.79 (-18.94) and $z(t)$=-4.18 (-3.58); for the OECD: $z(\rho)$=-33.82 and $z(t)$=-4.81. Hence, both time series of the change rate of labour force are stationary or I(0) processes.

There are three time series associated with unemployment in Austria shown in Figure 4, as defined by national statistics approach (AMS), Eurostat, and the OECD. It is illustrative to trace the definitional changes. Currently, the OECD and Eurostat use very similar approaches. There was a period between 1977 and 1983 when the OECD adopted the national definition, which was different from the one used by Eurostat. During a short period between 1973 and 1977, all series were very close. Major changes occurred between 1982 and 1987. The unemployment curves in Figure 4 are characterized by two distinct branches: a low (~2%) unemployment period between 1960 and 1982 and a period of higher unemployment (~4% for the OECD and Eurostat, and ~6.5% for the AMS) since 1983. These switches between various definitions present additional obstacles to obtaining a unique relationship between labour force and unemployment. The AMS definition is based on administrative records and might be the most consistent among the three, However, this definition differs from the one recommended by the International Labour Organization. Since all unemployment time series contain two distinct segments with a big step between them it is hardly stationary. Formally, all tests did not reject the null of a unit root due to the artificial structural break between 1982 and 1987. When corrected for this step, the rate of unemployment should be stationary.

The above discussion explains why one cannot model unemployment over the whole period by a unique linear relationship. There was a period of substantial changes in units of measurement between 1982 and 1987. Therefore, we model the rate of unemployment during the periods before 1982 and after 1986 separately. The period between 1982 and 1987 is hardly to be matched by a linear relationship. Results of the modelling are presented in Figure 5, where the AMS unemployment curve is matched by the following relationships:

$u(t) = 0.35l(t) + 0.0260$; *t<1982*

$u(t) = 0.70l(t) + 0.0705$; *t>1986*     (8)

The NAC labour force without a time lag is used to predict unemployment using equation (8). The absence of a lag might be presumed as natural behaviour of labour force and unemployment, as one of the labour force components. All empirical coefficients in (5) provide the best fit between the observed and predicted curves. From this Figure and relationship, one can conclude that there was a step change in the average level of unemployment from approximately 0.03 during the years before 1982 to 0.07 for the period after 1986. Accordingly, the slope doubled in 1987 indicating higher sensitivity of unemployment to the change in labour force under new definitions introduced between 1982 and 1987.

The estimate of root-mean-square error (RMSE) obtained for the OECD unemployment time series in Austria during the period between 1983 and 2012 is 0.0054. Therefore, the inherent variation in this data series is extremely low and thus difficult to model. Not surprisingly, the standard deviation for the model error in (8) is 0.0055. This relationship is accurate but unreliable.

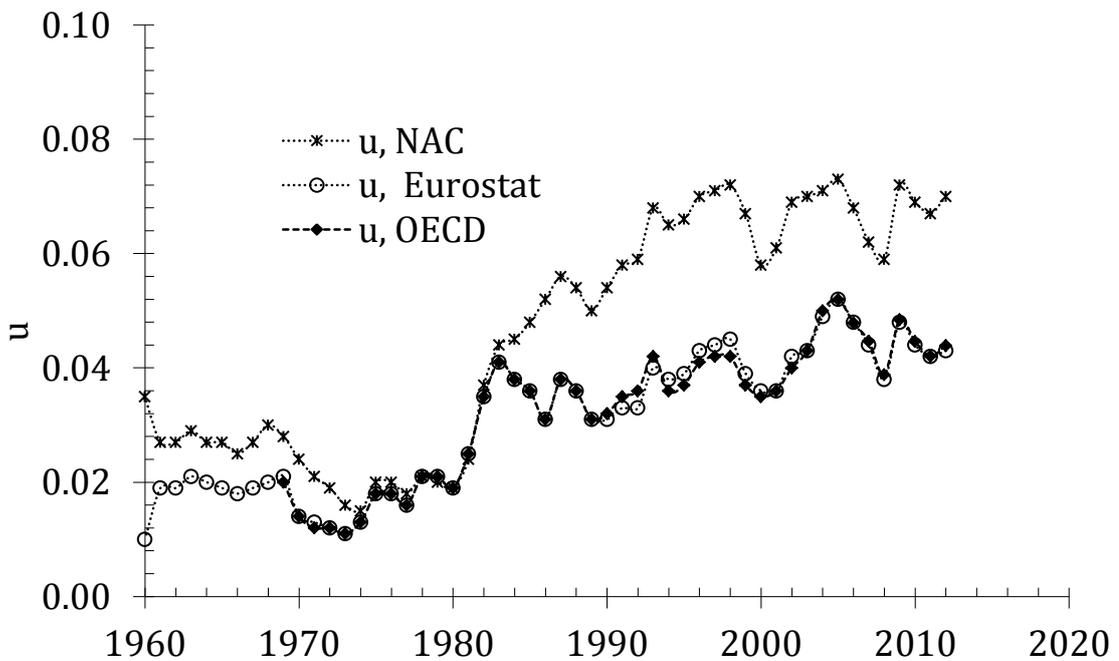

Figure 4. Estimates of unemployment rate in Austria according to definitions given by the AMS, Eurostat, and the OECD.

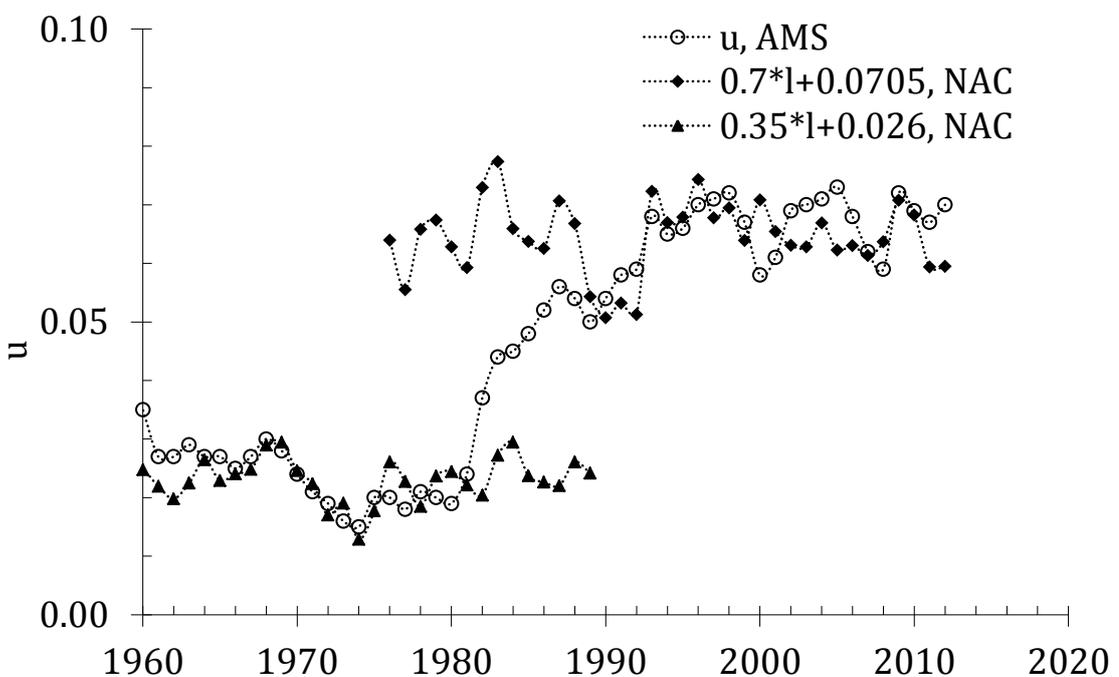

Figure 5. Comparison of the observed (AMS) unemployment rate and that predicted from the NAC (AMS+HSV) labour force. The changes in the unemployment and labour force definitions between 1983 and 1987 make it impossible to fit the unemployment curve during this period.

Figure 6 depicts the observed and predicted, annual and cumulative, rates of price inflation in Austria for the period between 1960 and 2012. The inflation time series is represented by the GDP deflator determined by Eurostat: all prices are converted in Euro before 2001. As mentioned above, there was a significant definitional and procedural change in the labour force (employment and unemployment separately) surveys in the 1980s. To compensate the effect of the artificial change in measurement units we introduced a structural break in our model. Two periods are described by two different linear relationships between the dependent and independent variables. The labour force is taken according to the AMS definition. Two relationships predicting inflation are as follows:

$$\pi(t) = 3.846 l(t-0) + 0.0484; \quad 1965 \leq t \leq 1986$$
$$\quad (0.67) \quad\quad (0.009)$$

$$\pi(t) = 2.383 l(t-0) + 0.00021; \quad t \geq 1987 \quad\quad (9)$$
$$\quad (0.33) \quad\quad (0.004)$$

The estimates of coefficients and lag (0 years) in (9) were obtained altogether by the Boundary Element Method (BEM) with the least-square fitting between the integral curves over the entire period, with 1986 being the point of structural break. The break point has been also estimated by the fitting procedure. If the model residual for the integral variables is an I(0) process then they are cointegrated and the annual model error is an I(-1) process. All estimated slopes are reliable with p-values less than $10^{-3}$. Since all readings in both time series in (9) have non-zero uncertainties the linear regression technique would underestimate the slopes. If to use the coefficients obtained by linear regression, the cumulative curves in Figure 6 would diverge. By design, the BEM guarantees perfect convergence between integral solutions, and thus, between the annual measurements.

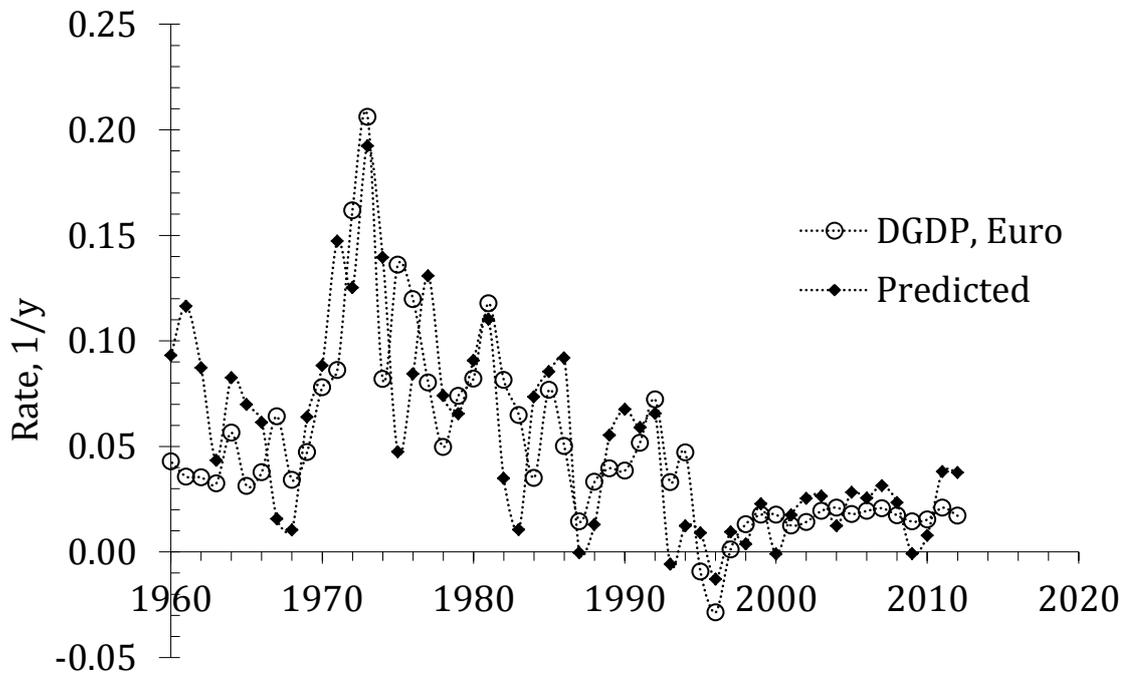

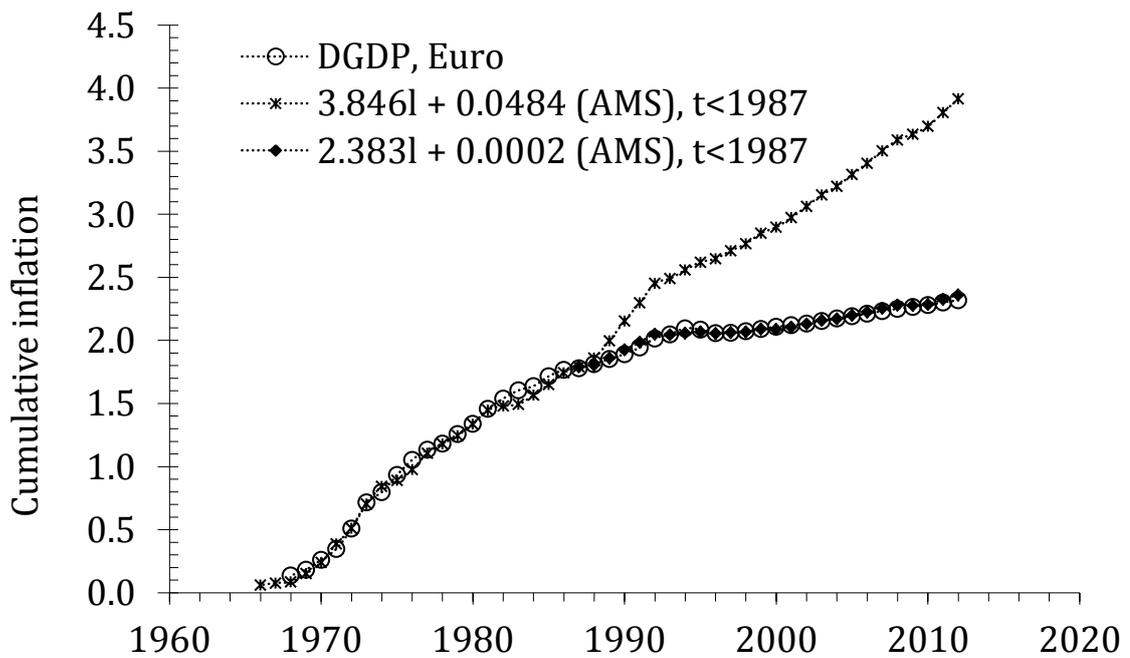

Figure 6. Comparison of the observed (DGDP, Euro) and predicted inflation in Austria. The upper panel displays annual readings and the lower one – cumulative inflation since 1965. The periods before and after 1986 are described separately. In order to demonstrate the effect of structural break the earlier relationship is extended beyond 1986.

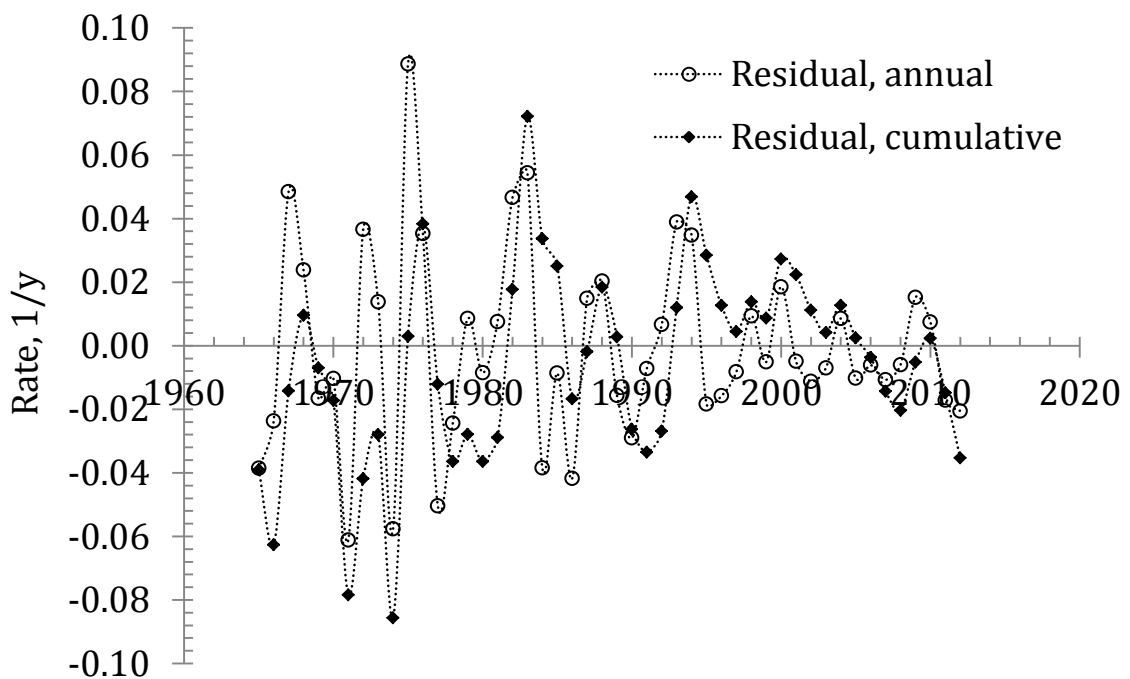

Figure 7. Model errors for the annual and cumulative curves in Figure 6. For (9), the BEM seeks to minimize the RMS cumulative error over the entire period. If the model error for integral variables is an I(0) process then the annual model error is an I(-1) process.

In the original DGDP Euro time series, the average values is 0.049 and standard deviation σ=0.044. The model residual (see Figure 7) is characterized by RMSE=0.026. The naive forecasting technique (AR(1) model) provides RMFSE=0.032 for the residual time series, i.e. the first difference of the original DGDP series. Therefore, the model based on the change in labour force without autoregressive terms definitely outperforms the naive forecasting for the whole period. (We assume that the level of labour force can be accurately estimated one year ahead.) Our model retains the same functional (linear) link between the studied variables, which is perfectly parsimonious.

Relations (9) also outperform the naive forecasts in both periods taken separately. After 1986, RMSE=0.014 which is lower than 0.019 for the AR(1) model. The DGDP time series is characterized by $\sigma$=0.020 for this period. Between 1965 and 1986, the estimated standard deviations are 0.036, 0.043, and 0.043, respectively. The coefficient of determination $R^2$=0.66 for the entire period, i.e. the change in labour force explains 66 % of the DGDP variability. For the cumulative curves in Figure 6, RMSE=0.030 (marginally higher than for the annual model error), $R^2$=0.999.

Since the cumulative curves are both non-stationary, I(1), processes we tested them for cointegration. The Johansen test showed the maximum rank 1, i.e. the existence of one long-term equilibrium relation between the cumulative curves. (The trace statistic is 0.214* with the 5% critical value 3.84.) The null hypothesis of a unit root in the residual time series (the difference between the cumulative curves) was rejected by the Augmented Dickey-Fuller test (-3.74, with 1% critical value of -3.60). The Phillips-Perron test has also rejected the null of a unit root: $z(\rho)$=-20.91 (-18.70) and $z(t)$=-3.67 (-3.60). Formally, the observed and predicted cumulative time series are cointegrated, and thus, the extremely high coefficient of determination is not biased.

With the unbiased coefficient of determination of 0.999, one can interpret the relation between the cumulative curves as a causal link. The level of labour force is controlled by the size of working age population and the rate of participation in labour force. The former variable depends on the integral value of births 15 and more years ago, total deaths in working age population, and net migration. All these processes are far beyond the influence of price inflation. The rate of participation in labour force is rather a slow evolving variable and depends on economic activity. Hence, the cumulative price inflation cannot affect the net change in labour force, i.e. the cumulative change in labour force in (9). On the other hand, the change in labour force explains 99.9% of variability in the cumulative inflation. One can substitute the price change with labour change during the same period.

The slope ratio in (9) is 3.84/2.38 ~ 1.61 and the intercept dropped from 0.0484 to zero. The change in the slopes is consistent with the changes in definition of labour force between 1982 and 1987 – gradually more and more persons were counted in as employed and unemployed with a substantial increase in the labour force level. This increase resulted in the growth of annual increments and the decrease of the slope (or sensitivity) in (9). Therefore, the inflation sensitivity to the new measures of labour force, or new units of measurement, in Austria decreased in 1987. When the relationship obtained for the first period is used for the second period, the rate of inflation is significantly overestimated, as shown in the lower panel of Figure 6. The deviation between the two predicted curves after 1986 clearly demonstrates the influence of the changes in measurement definitions for quantitative modelling of economic parameters.

The predicted curves are in good agreement with actual inflation. A prominent feature is the overall similarity between 1968 and 1975, when the highest changes in the inflation rate were observed: from 0.034 in 1968 to 0.21 (21% per year) in 1973, and back to 0.05 in 1978. Overall, the Phillips curve, the NKPC or any other model that rely on autoregressive properties of inflation fail to describe this kind of dynamics. Due to the LSQ fitting, the model coefficients and lags are adjusted to the biggest changes, i.e. to the peak in 1973. However, Figure 1 shows that two other measures of inflation have peak values at different times.

Our model fully relies on labour force estimates. Several simple measures have been proposed in order to improve the quality of labour force measurements and to obtain more reliable statistical estimates. Due to lack of information on quantitative characteristics of the revisions applied to the Austrian labour force series we cannot explicitly correct for possible step revisions. In the absence of correcting measurements, it is natural to apply the moving average technique. Three-year moving average, MA(3), suppresses the noise associated with the labour force measurements and also removes the half-year shift in timing between inflation and labour force readings. In the lower panel of Figure 8, the MA(3) smoothed predicted and observed curves are shown in order to demonstrate the accuracy of the predictions when longer bases are used to estimate inflation and labour force. The only period of larger deviations between these curves is from 1984 to 1986, i.e. near the structural break.

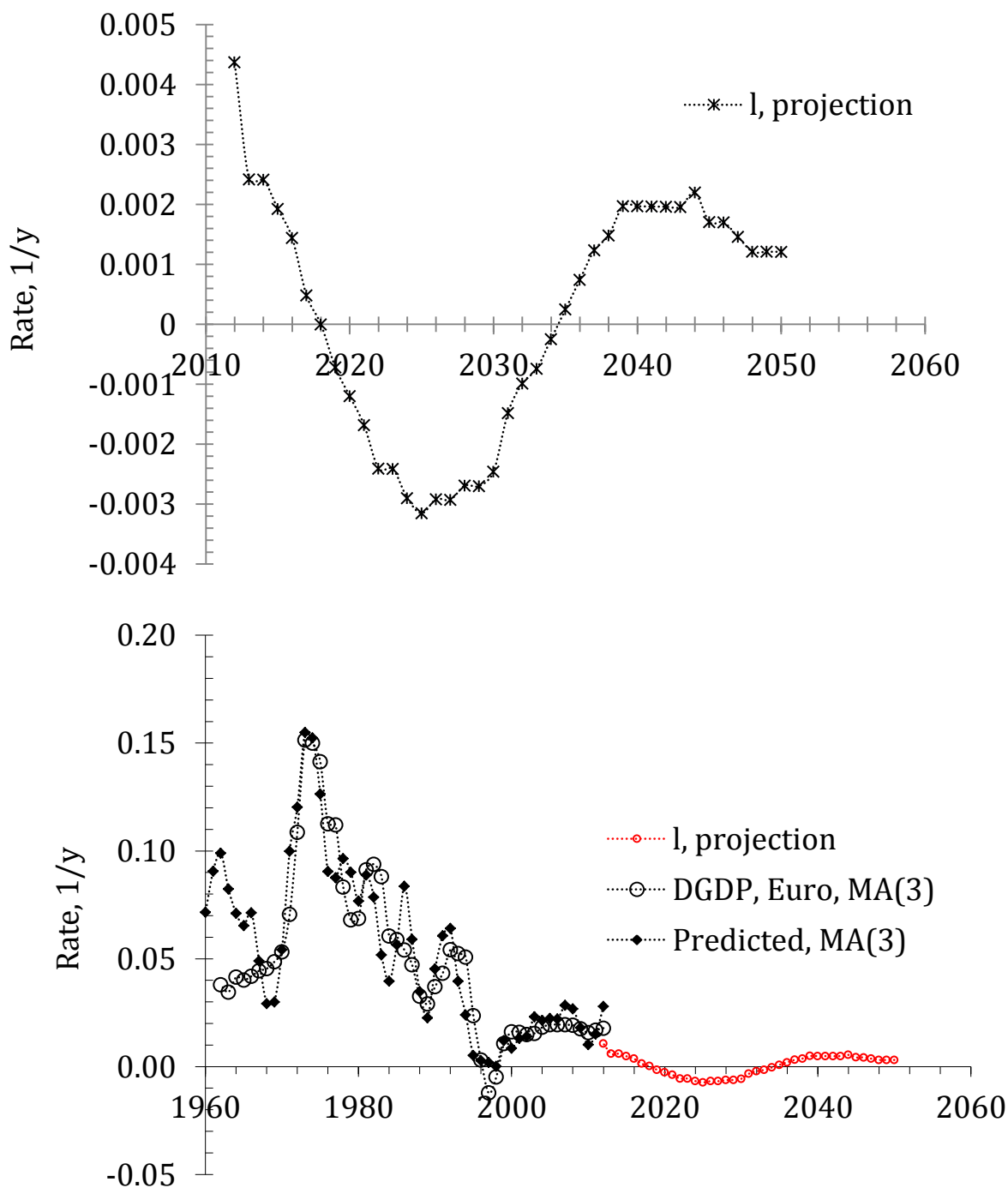

Figure 8. Upper panel: Labour force projection borrowed from Statistik Austria. The period from 2018 to 2034 is characterized by a negative growth rate. Lower panel: Inflation projection obtained from the labour force projection according to (9). The measured and predicted curves between 1960 and 2012 are smoothed with MA(3) in order to demonstrate the accuracy of the predictions when longer bases are used to estimate inflation and labour force.

The upper panel of Figure 8 depicts the change rate of labour force from 2011 to 2050, as obtained from a projection published by Statistik Austria (2013). Not considering the accuracy and reliability of the 40-year-ahead projection we focus on the overall behaviour of inflation during the same period. In the lower panel, the GDP deflator projection obtained by (6) from the labour force projection. Since the intercept in (9) after 1986 is practically zero, the negative rate of labour force is equivalent to deflation. Therefore, from 2018 to 2034 a deflationary period is expected as related to the GDP deflator.

Figures 9 and 10 show the results of a similar analysis for two different measures of inflation: the NAC CPI and GDP deflator. From Figure 1, one can conclude that both measures are close, but diverge during extended periods and have peaks at different years. Therefore, the CPI (7) and DGDP NAC (8) are slightly different:

$$\pi(t) = 1.755(t\text{-}2) + 0.0361; \quad 1965 \leq t \leq 1986$$
$$\quad (0.30) \quad\quad (0.004)$$

$$\pi(t) = 0.958l(t\text{-}2) + 0.0130; \quad t \geq 1987 \quad\quad (10)$$
$$\quad (0.17) \quad\quad (0.002)$$

$$\pi(t) = 1.734l(t\text{-}1) + 0.0373; \quad 1965 \leq t \leq 1986$$
$$\quad (0.23) \quad\quad (0.003)$$

$$\pi(t) = 1.439l(t\text{-}1) + 0.00479; \quad t \geq 1987 \quad\quad (11)$$
$$\quad (0.17) \quad\quad (0.002)$$

The most important difference is the nonzero lags, $t_1$, in (10) and (11) before and after the structural break. The rate of CPI inflation lags behind the change in labour force by two years and the DGDP NAC – by one. This allows inflation forecasting at a two- and one-year horizons, respectively. To a large extent, these lags are defined by the peak inflation in the 1970s. Since the late 1990s, inflation is constant and hardly adds something to the resolution of time lags.

All coefficients are obtained using the BEM with the *LSQ* fitting between the cumulative curves. The slope ratio is 1.83 and 1.21 for the CPI and DGDP, respectively. Before 1986, both intercepts are close. After 1986, the DGDP intercept is close to zero and does not prevent deflation when the level of labour force falls. For the CPI, the long-term rate of inflation is 1.3% per year, which is slightly lower the target value of 2%.

The observed and predicted time series of the CPI and DGDP were successfully tested for cointegration. The Johansen test showed the maximum rank 1 for both cases. For the relevant residuals, the ADF and Phillips-Perron tests rejected the null of a unit root. Hence, the observed and predicted cumulative curves are cointegrated, with the predictor (the change in labour force) leading by one (DGDP) and two (CPI) years. The coefficient of determination for the cumulative curves in both cases is larger than 99.8%.

The CPI time series is characterized by the average value of 0.035 and standard deviation $\sigma$=0.021 for the period between 1965 and 2012. These values are close to those obtained for the NAC DGDP series: 0.034 and 0.022, respectively. Between 1965 and 2012, the naïve prediction for the CPI has RMSFE=0.014 at a one-year horizon and RMSFE=0.018 at a two-year horizon. For the DGDP, at a one-year horizon RMSFE=0.011. These values define the reference forecasting precision.

For the observed and predicted CPI NAC times series, the coefficient of determination, $R^2$=0.68, is lower and RMSFE=0.0118 (at a two year horizon) is higher than those for the DGDP NAC: $R^2$=0.82 and RMSFE=0.0095 (at a one year horizon), respectively. Even small differences between the DGDP and CPI (cross correlation coefficient is 0.92) result in a large difference in statistical estimates. However, both models are superior to the naive forecasting. The CPI forecast at a two-year horizon is by 70 per cent (0.011/0.018) more precise. The DGDP forecasting precision is by 20% better than the naive one. This result is obtained with the AMS labour force time series, which is likely noisy due to data scaling from the survey to the total population. The precision of labour force estimates can be further improved. It is worth noting that the estimates of RMS(F)E obtained for the cumulative time series are as follows: 0.027 for the DGDP Euro; 0.019 for the DGDP NAC; 0.031 for the CPI NAC.

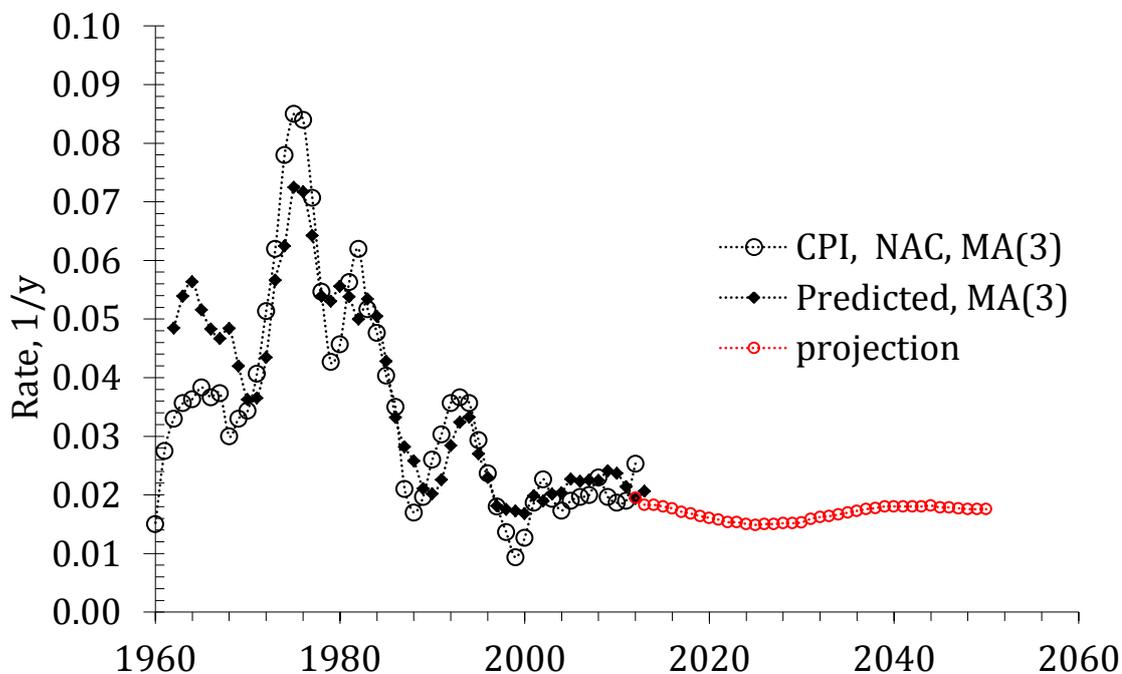

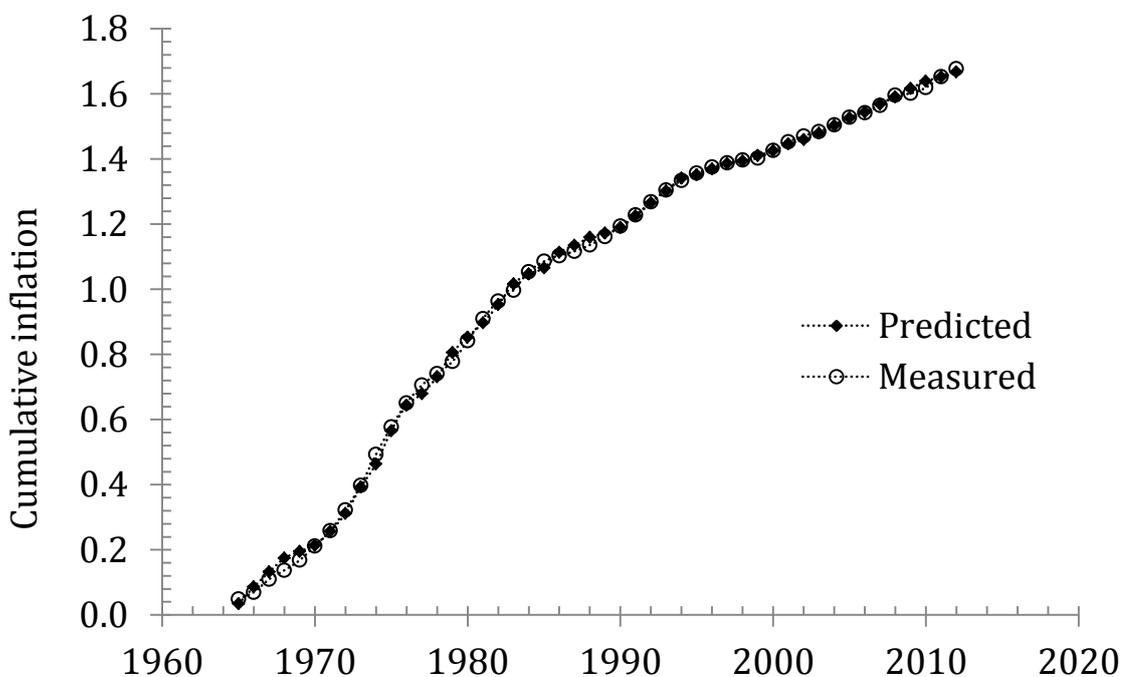

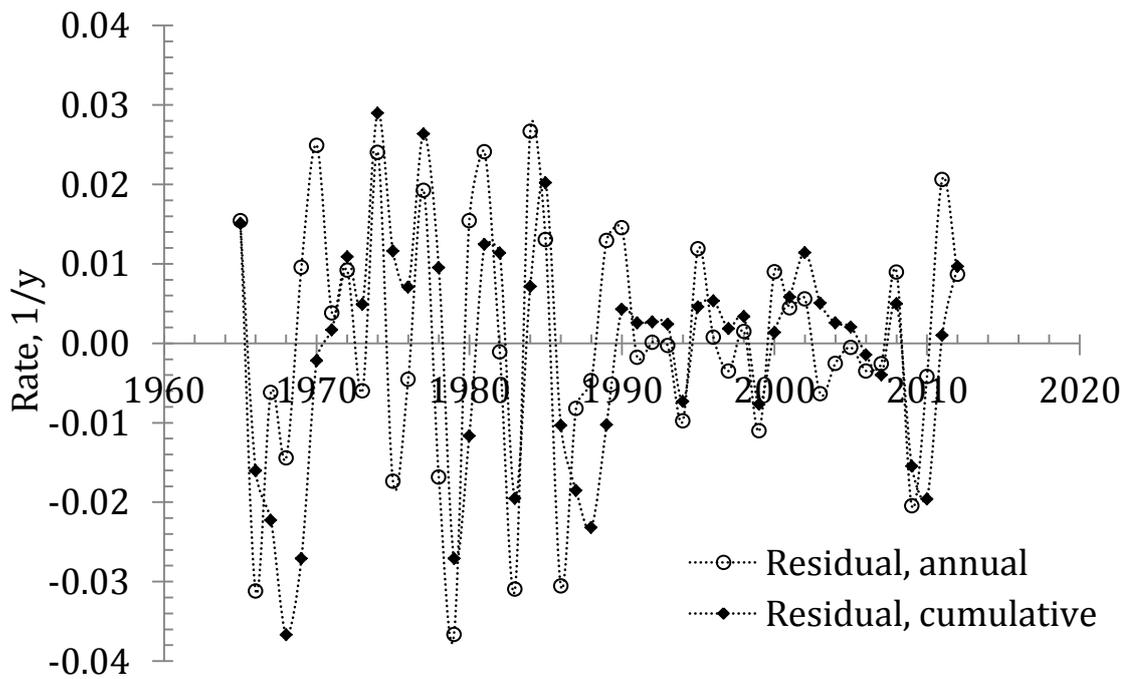

Figure 9. Comparison of the observed (CPI NAC) and predicted inflation in Austria. Upper panel: The annual readings of observed and predicted inflation smoothed with MA(3). The rate of inflation is projected through 2050. Middle panel: The cumulative curved of observed and predicted inflation since 1965. Lower panel: The annual and cumulative model errors.

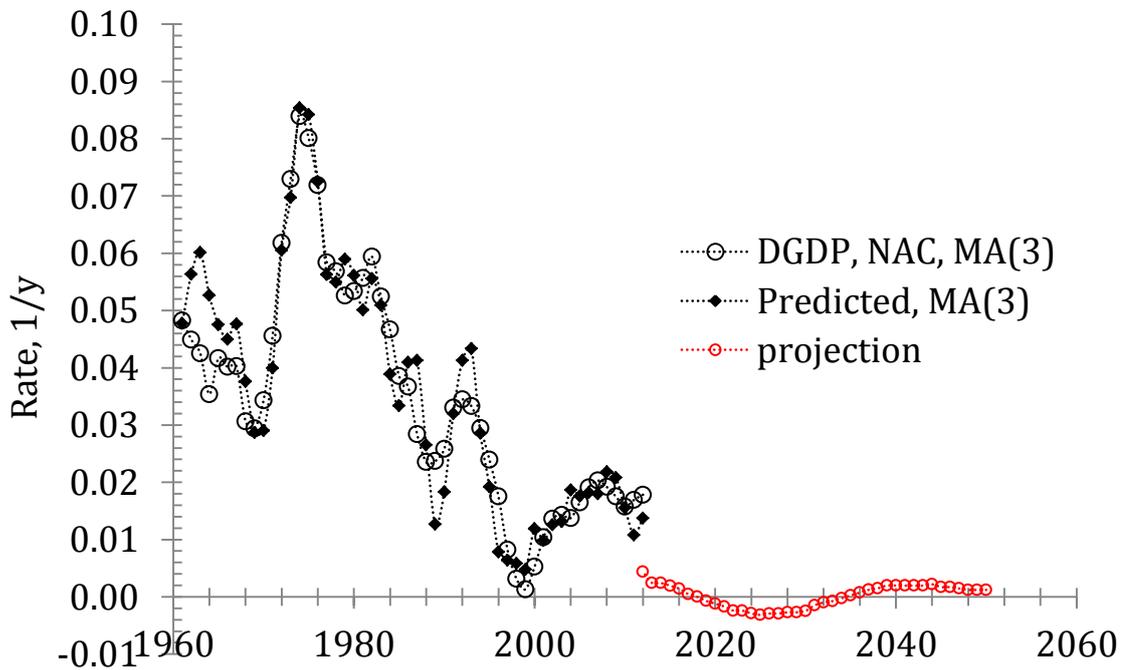

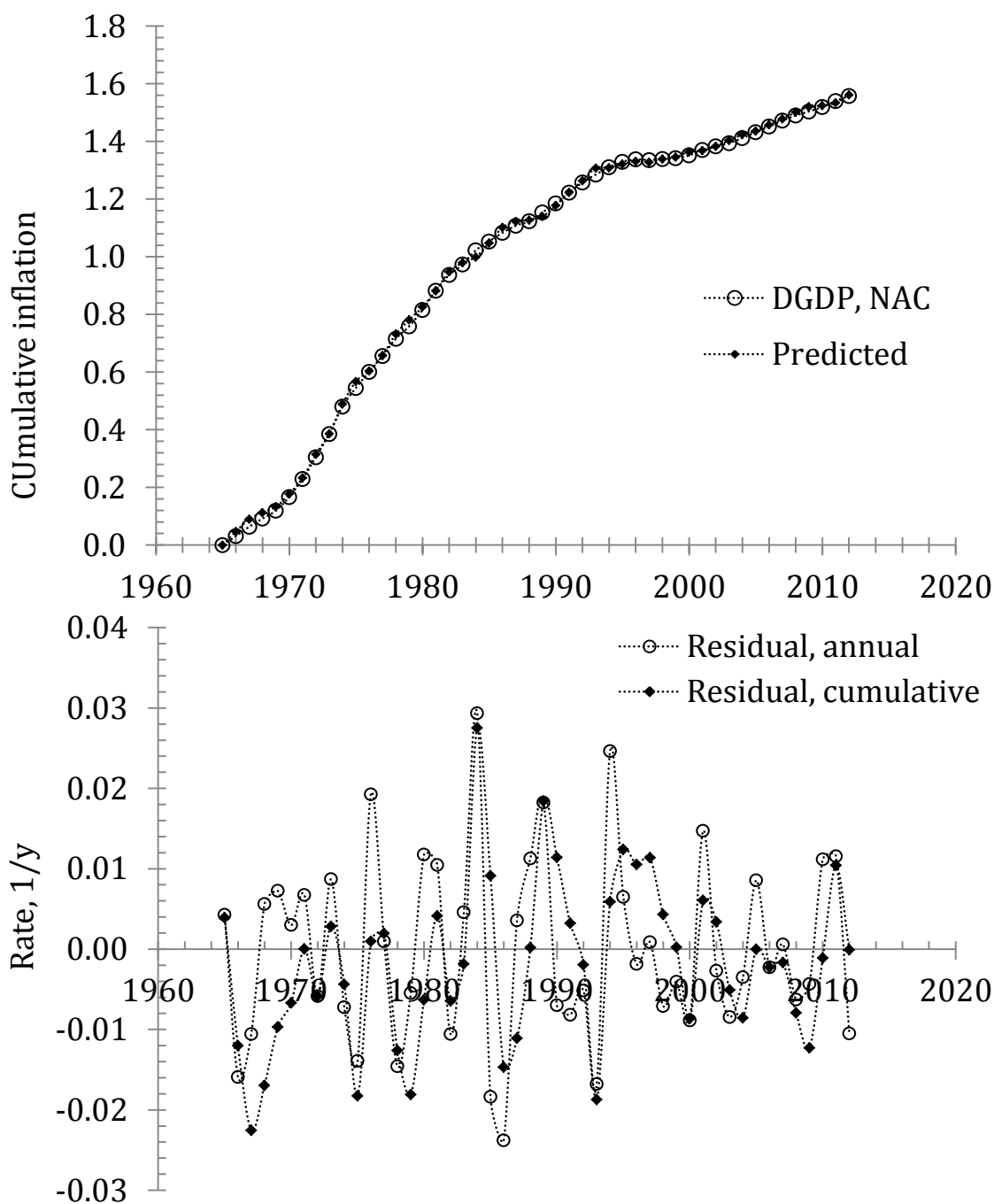

Figure 10. Same as in Figure 9 for the DGDP NAC.

The DGDP NAC, being the most complete measure of price inflation reflecting the behaviour of the Austrian economy as a whole, demonstrates the lowermost scattering in the model residual: RMSFE=0.0095 for the period between 1965 and 2012. Normalizing to the standard deviations in the original DGDP series one obtains 0.0095/0.022=0.43 for the NAC DGDP and 0.026/0.044=0.59 for the DGDP Euro. For the CPI, this ratio is 0.011/0.021=0.52. The relative improvement in the variability of residual errors does not differ much.

The labour force projection implies a very low but positive rate of growth in consumer prices through 2050. The DGDP NAC will experience approximately 15 year deflationary period in line with the prediction for the GDDP Euro. The consumer price inflation will hover around 1.3% per year between 2010 and 2050, as the intercept in (11) for the period after 1986 and the labour force projection imply. This prediction undermines the capability of the OeNB to retain inflation at the target level. In reality, the period of 2% inflation in Austria observed in the 2000s is the result of quasi-constant growth in labour force (see Figure 3). With the decaying labour force, the Austrian National Bank will not be able to return inflation to its target value.

It has been confirmed above that both inflation and unemployment in Austria are linear functions of labour force. For inflation, time lags depend on the selected measure. All relationships are characterized by a structural break. There is no need to apply generalized relationship (4). However, relationships (8) through (11) demonstrate an excellent predictive power separately and their sum should also work well. There is another important feature of the original time series associated with the use of (4). Measurement errors make the prediction of annual time series less reliable during the periods of weak changes (*e.g.*, the 2000s), when the annual change in labour force is lower than the uncertainty of the labour force estimates. Then, the observed change is statistically insignificant, as we have obtained for the rate of unemployment in Austria. Relationship (4) provides a potential way to improve the overall match. Since all the involved variables have almost independent measurement errors one can expect additional destructive interference of these errors when they are used altogether.

For the generalized representation, we model the DGDP NAC. We have estimated coefficients in relationship (4), with a possible break between 1980 and 1990, using the BEM and the *LSQ* fitting applied to the OECD unemployment rate and the AMS labour force as predictors:

$$\pi(t) = 1.0l(t) - u(t) + 0.068; \quad t \leq 1986$$

$$\pi(t) = 0.8l(t) - u(t) + 0.077; \quad t \geq 1987 \tag{12}$$

Figure 11 displays the DGDP NAC and inflation predicted by (12). We expect that (12) is less sensitive to the changes in unemployment and labour force definitions. Unemployment is a smaller share of labour force and any change in the rate of unemployment is automatically included into the labour force change. Nevertheless, the changes in unemployment and employment definitions are not necessarily synchronized. Despite these definitional problems, the agreement between the predicted and observed curves is remarkable. For the annual readings between 1962 and 2012, $R^2$=0.76 and RMSE=0.010. This is a better result than that obtained from the naive prediction (RMSFE=0.011) for the same period, but it is inferior to the RMSFE provided by (11). The cumulative curves are cointegrated and $R^2$=0.999. All in all, the labour force and unemployment describe the integral evolution of inflation since 1962 with an increasing relative accuracy.

**Conclusion**

On average, the rate of CPI inflation in the 21$^{st}$ century was 2.1% per year and only 1.7% per year for the GDP deflator. The CPI inflation is close to that explicitly defined by the monetary policy adopted by the European System of Central Banks and by the National Bank of Austria (OeNB, 2013). This observation is considered as the result of OeNB's monetary policy. At the same time, the rate of the overall price inflation has been following the revealed dependence on the change in labour force. Hence, the monetary policy oriented to price stability is currently synchronized with the natural economic evolution. This monetary policy has not been disturbing (i.e. showing any effect on) relationship (11) describing the last 50 years of inflation. Therefore, the future evolution of price inflation will be likely controlled by the change in labour force not by the OeNB. The projection of workforce published by Statistik Austria implies that in the long run the rate of CPI inflation will be 1.3% per year and the GDP deflator will sink below zero between 2018 and 2034.

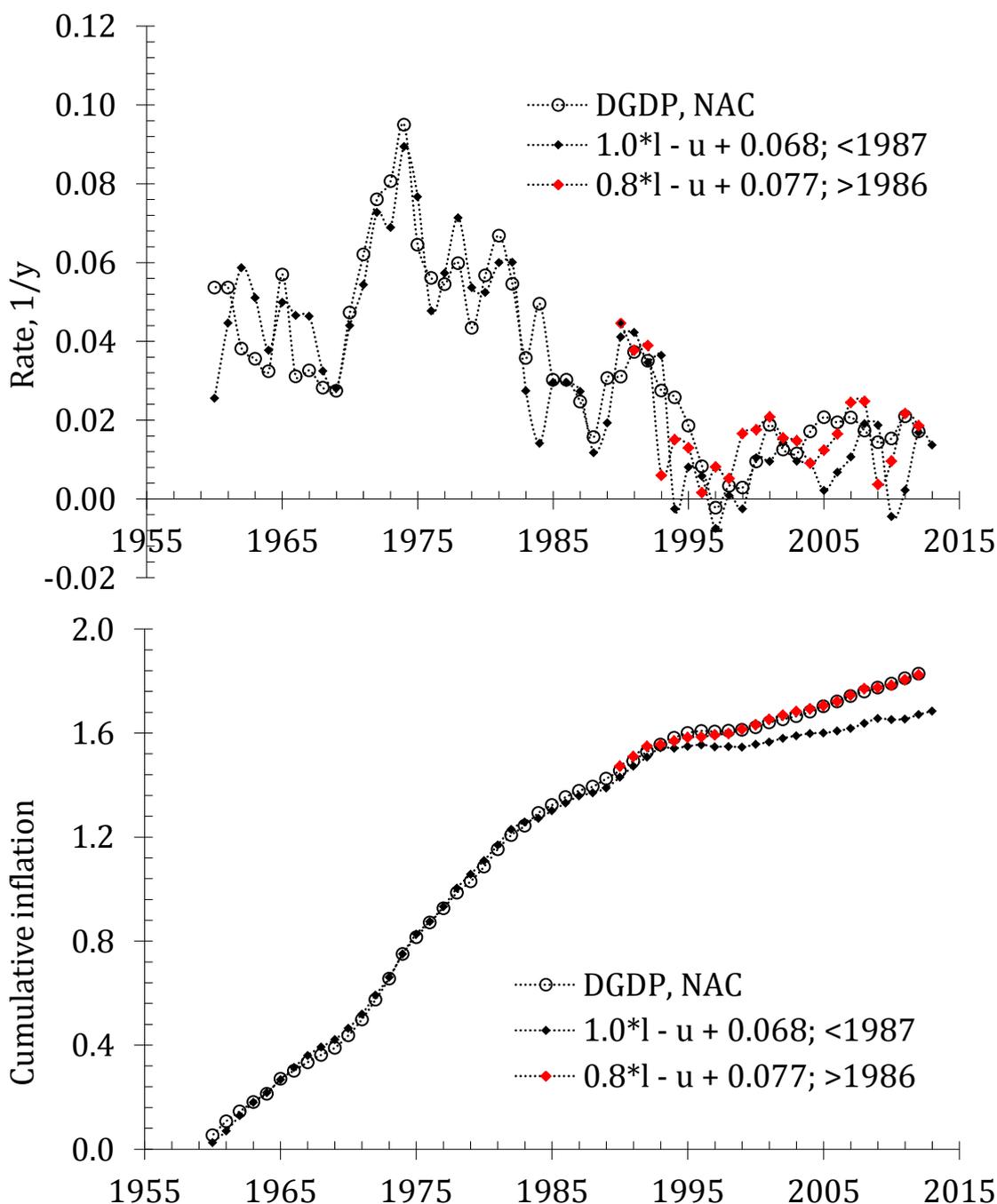

Figure 11. Comparison of the observed (DGDP Euro and predicted inflation in Austria. The upper panel displays annual readings and the lower one – cumulative inflation since 1962. The predicted inflation is a linear function of the labour force change and unemployment as defined by (12).

Austria provides a good opportunity not only to model the dependence between inflation, unemployment, and labour force change, but also to evaluate the consistency of various definitions of the studied variables. Despite the well-documented changes in units of measurements, these variables do not lose their intrinsic links. There is no reason to think that the inherent relationships will disappear in the future.